\documentclass{iopart}
\usepackage{graphicx,epsf,iopams}

\begin{document}
\title{Fluctuation relations and coarse-graining}

\author{Saar Rahav$^1$, Christopher Jarzynski$^{1,2}$}
\address{$^1$ Department of Chemistry and Biochemistry, University of Maryland, College Park 20742, USA.}
\address{$^2$ Institute for Physical Science and Technology, University of Maryland, College Park 20742, USA.}

\date{\today}
\begin{abstract}
We consider the application of fluctuation relations to the dynamics
of coarse-grained systems, as might arise in a hypothetical experiment
in which a system is monitored with a low-resolution measuring apparatus.
We analyze a stochastic, Markovian jump process with a specific structure
that lends itself naturally to coarse-graining.
A perturbative analysis yields a reduced stochastic jump process
that approximates the coarse-grained dynamics of the original system.
This leads to a non-trivial fluctuation relation that is approximately
satisfied by the coarse-grained dynamics.
We illustrate our results by computing the large deviations of
a particular stochastic jump process.
Our results highlight the possibility that observed deviations from
fluctuation relations might be due to the presence of unobserved
degrees of freedom.
\end{abstract}

\maketitle

\section{Introduction}
\label{intro}

Coarse-graining is at the heart of equilibrium statistical
mechanics. It is used to describe systems with a macroscopic number
of degrees of freedom with the help of a much smaller number of
variables. A familiar example is the description of a gas in terms of
its slowly varying local density rather than the position of all its molecules~\cite{Chaikinbook}.
Coarse-graining is particularly important in the study of
phase transitions, and more generally, interacting systems. It
is a crucial building block in renormalization group
theory~\cite{Amitbook,Fisher1998}. Even far from a phase transition
the combination of coarse-graining with the identification of
relevant and irrelevant variables can sometimes be used to replace
seemingly complicated interactions by (approximate) simpler ones.

The theory of systems out of equilibrium is less developed and, in
particular, the possible uses of coarse-graining in such a theory
are not as clear. One approach, recently reviewed by
Derrida~\cite{Derrida2007}, uses large deviation theory to address
the fluctuations of a local density or some other coarse-grained field.
This leads to a functional of the local density that characterizes the steady
state and fluctuations around it.

In recent years, a different approach has been employed to study
systems far from equilibrium. In this approach one investigates
symmetries, related to time-reversal, that are associated with the
probabilities of observing sequences of events.
Such investigations have led to the discovery of a number of interesting
results, collectively known as {\it fluctuation relations},
including some that provide information related to the likelihood to
observe so-called ``second law violations''~\cite{Evans1993,Evans1994,Gallavotti1995}.
Loosely speaking, fluctuation relations have the form
\begin{equation}
\label{eq:ft_intro}
\ln \left[ \frac{P (\Delta S = q T)}{P (\Delta S = -q T)}\right] = q
T + o(T),
\end{equation}
where $P(\Delta S)$ is the probability to observe a change $\Delta
S$ in entropy, $T$ is the observation time, and $q$ is the observed
entropy creation rate. [The notation $o(T)$ indicates a correction
that grows more slowly than $T$, that is $\lim_{T\rightarrow\infty}
o(T)/T = 0$.] These results are quite general, they remain valid far
from thermal equilibrium, and they are part of an active field of
research focused on the application of the laws of thermodynamics to
microscopic systems~\cite{Bustamante2005}.

Fluctuation relations are derived by considering statistical
distributions of microscopic trajectories of the system of interest.
For any such trajectory, we assign a value to a quantity $\Delta S$,
which we interpret physically as a measure of entropy production. By
comparing the probability of observing a given trajectory ($\gamma$)
with that of its time-reversed counterpart ($\bar\gamma$), the
symmetry represented by equation~(\ref{eq:ft_intro}) emerges (see
section~\ref{fluct}). This approach, however, runs into difficulties
when coarse-graining is involved. Consider an experiment in which
some degrees of freedom cannot be observed, perhaps because they
evolve on time scales much faster than the response time of the
detector. In this case our observation of the system gives us a {\it
coarse-grained} -- i.e.\ a ``smeared'', or locally averaged --
trajectory, rather than a full microscopic record of the evolving
state of the system. It is not at all evident that such
coarse-grained trajectories satisfy the same symmetry relations as
their fully microscopic counterparts. Moreover, important
information might be lost due to coarse graining; specifically, the
expression for entropy production, $\Delta S$, might contain
contributions from those degrees of freedom that are not resolved by
the detector. Under these circumstances, can we define some sort of
``coarse-grained entropy production'', $\Delta S^{\rm CG}$, which
satisfies a fluctuation relation?

Our goal in this paper is to give partial answers to
such questions. We will examine a simple system for which an assumed separation of
time scales leads to a simple coarse-graining transformation.
The system we study is a stochastic jump process whose probability
distribution satisfies a Master equation.
Fluctuation relations for such jump processes were first derived
by Lebowitz and Spohn~\cite{Lebowitz1999}, following Kurchan's
analysis of Langevin processes~\cite{Kurchan1998}.
These two papers laid the groundwork for much subsequent work on
stochastic fluctuation relations,
which has recently been reviewed by Harris and
Sch\"utz~\cite{Harris2007}. In the present paper,
we impose a separation of time scales and argue that, when the
slow component of the dynamics is accurately measured, a non-trivial
fluctuation relation, which differs from the one valid for the
microscopic system, is (approximately) satisfied. This relation
involves a coarse-grained entropy production, which
depends only on the slow component of the evolution.

We note that in the context of fluctuation theorems, an explicit separation
of time scales has appeared as well in the work of Zamponi {\em et.al.}~\cite{Zamponi2005}.
These authors have studied a Langevin particle interacting with
a nonequilibrium thermal environment consisting of slow and fast components
at different temperatures (see also \cite{Ilg2006}).
For such a system a fluctuation-dissipation relation
between spontaneous and externally induced fluctuations can be used
to define an {\it effective temperature}.
The system was shown to satisfy a generalized fluctuation theorem
formulated in terms of this effective temperature.

In section \ref{coarse} we will present the model used and apply a
coarse-graining transformation, replacing it by a simpler system.
The (approximate) coarse-grained system is also Markovian and
therefore satisfies a fluctuation relation of its own. In
section~\ref{fluct} we use the existence of a coarse-grained
counterpart to define coarse-grained trajectories, and a
coarse-grained entropy production $\Delta S^{\rm CG}$. These
trajectories can be heuristically viewed as measurements obtained
with a limited-resolution detector. We show that the coarse-grained
entropy production approximately satisfies a fluctuation relation
(for the original system). A simple example is numerically studied
in section \ref{example}, in order to illuminate the results
obtained in previous sections. The results are summarized in section
\ref{disc}.

\section{Stochastic systems with simple coarse-graining}
\label{coarse}

While there are many physically relevant and interesting examples of systems
with natural coarse-grained counterparts,
we will focus on stochastic jump processes governed by master equations.
After briefly
reviewing general properties of such processes, we present models with
special structure, which makes them amenable to coarse-graining.
We then use simple perturbation theory to motivate a coarse-graining transformation
that replaces the system of interest by a smaller, reduced system.

Consider a system composed of a finite number of distinct states.
The system can make transitions between these states, and these transitions
are assumed to be described by a Markov process.
Specifically, given two states $\sigma$ and $\sigma^\prime$,
the rate $R(\sigma | \sigma^\prime)$ is the probability per unit time
to make a transition to $\sigma$, from a current state $\sigma^\prime$.
While not all transitions are allowed (some of the rates vanish), we do
assume that if the transition from $\sigma^\prime$ to $\sigma$ is
allowed [$R (\sigma | \sigma^\prime) \neq 0$] then also the reverse
transition is allowed [$R (\sigma^\prime | \sigma) \neq 0$]. This
assumption is analogous, but not equivalent, to microscopic reversibility for
deterministic systems.

\begin{figure}[t]
\center{\includegraphics[scale=0.8]{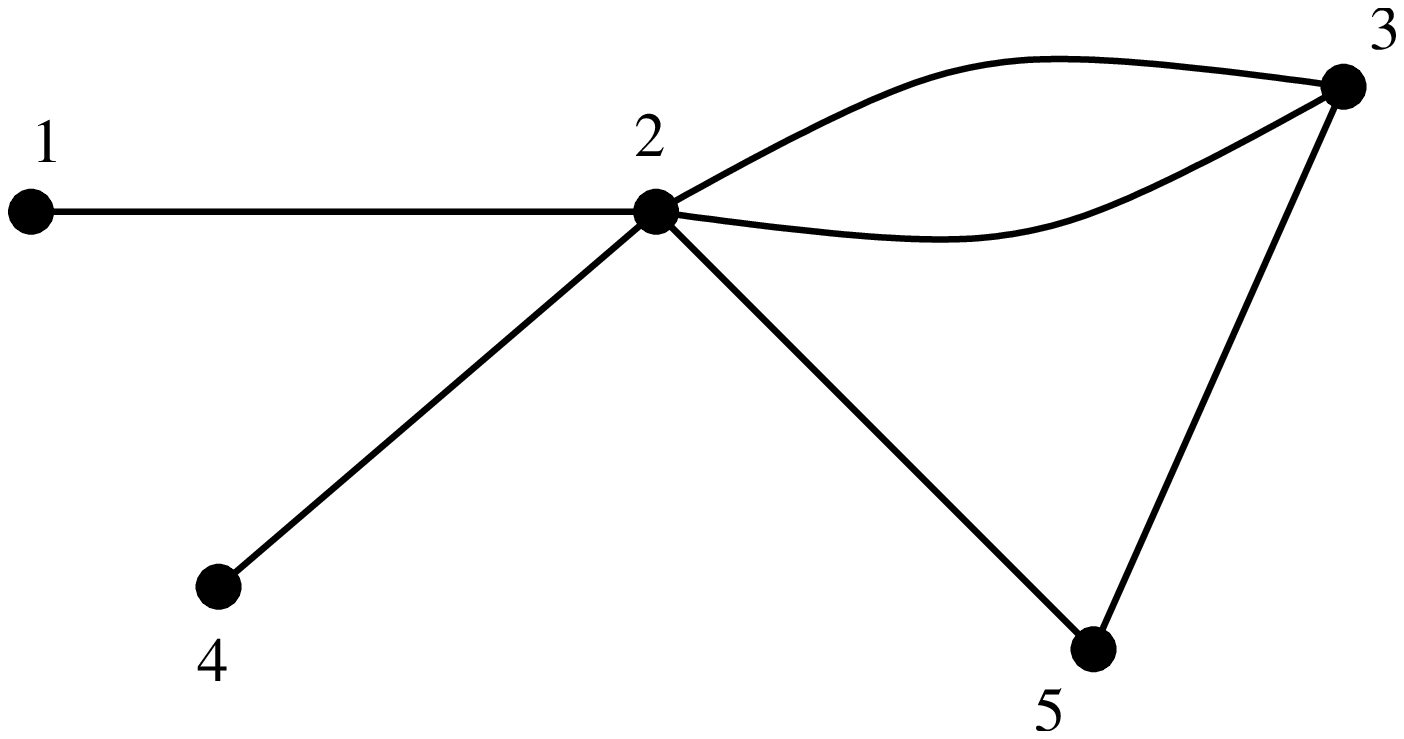}}
\caption{\label{examplegraph} A stochastic system described with the
help of a graph. The nodes correspond to the states of the system.
Each link represents a non-vanishing transition between two states,
and its reversed counterpart. The system is fully specified when the
transition rates are known.}
\end{figure}

It is useful to depict such a system with the help of a graph, in
which each node represents a state and each link represents a pair
of non-vanishing transitions. For instance, the graph shown in
figure~\ref{examplegraph}
illustrates a five-state system with six non-vanishing
links. Note that we allow multiple links between a given pair of states,
as between states 2 and 3.
In other words, there may exist
information that allows one to distinguish between
different ways to make a transition.
This implies that four rates -- namely,
$R_{\alpha=1,2} (3 | 2)$ and $R_{\alpha=1,2} (2 | 3)$ --
are needed to describe all transitions between states $2$ and $3$.

Such a process satisfies a master equation
\begin{equation}
\label{eqmaster} \frac{d  P (\sigma , t)}{d t}= \sum_{\sigma^\prime}
R (\sigma | \sigma^\prime) P (\sigma^\prime , t),
\end{equation}
where $P (\sigma , t)$ denotes the normalized probability to find the system in
the state $\sigma$ at time $t$.
It is useful to think of $P(t)$ as a vector whose components are
labeled by the index $\sigma$, and ${\bf R}$ as a matrix.
A non-diagonal element
$R (\sigma | \sigma^\prime)$ is given by the sum of the rates
from $\sigma^\prime$ to $\sigma$,
e.g.\ $R(\sigma | \sigma^\prime) = \sum_\alpha R_{\alpha} (\sigma
| \sigma^\prime)$,
whereas a diagonal element $R(\sigma | \sigma)$ specifies the net probability rate
of transitions out of the state $\sigma$.
To ensure conservation of probability, the rates satisfy
\begin{equation}
\label{condition} \sum_{\sigma} R (\sigma | \sigma^\prime) =0.
\end{equation}
We assume that the graph is {\it connected}, that is, it is possible
to go from any state to any other state using the links. In that
case, the solution of equation (\ref{eqmaster}) decays with time to
a unique stationary solution $P^s$ satisfying~\cite{VanKampenbook}
\begin{equation}
\sum_{\sigma^\prime} R (\sigma | \sigma^\prime)  P^{\rm s}
(\sigma^\prime)=0.
\end{equation}
The stationary probability current $ J^s_\alpha
(\sigma^\prime,\sigma) = R_\alpha (\sigma | \sigma^\prime)P^{\rm s}
(\sigma^\prime) -R_\alpha (\sigma^\prime | \sigma) P^{\rm s}
(\sigma) $ measures the net flow from $\sigma^\prime$ to $\sigma$,
via the transition $\alpha$, in the stationary state. If
$J^s_\alpha=0$ for every transition, then {\it detailed balance} is
satisfied, and it is natural to think of the system as being in a
state of thermal equilibrium. For a generic transition matrix ${\bf
R}$, however, the currents $J^s_\alpha$ are non-vanishing, and then
we consider the system to be in an
    out-of-equilibrium steady state~\cite{Zia2006}.

\begin{figure}[t]
\center{\includegraphics[scale=0.5]{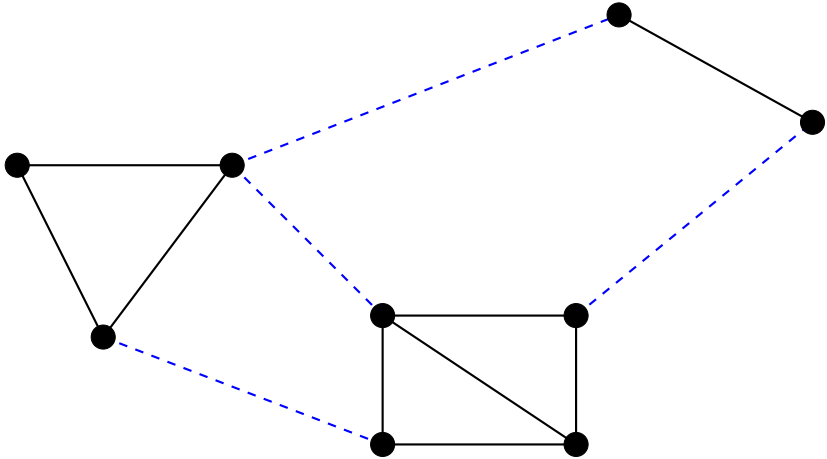}}
\caption{\label{bforecoarse} Graph of a stochastic system with a
meaningful coarse-graining transformation. The solid lines denote
links with large transition rates (of order unity). The dotted lines
denote weakly connected states (with rates of order $\varepsilon$).}
\end{figure}

Systems described by master equations are, generally, relatively
easy to handle mathematically. However, our interest is in systems
for which it is useful to coarse-grain the dynamics. This leads us
to examine systems with special structure. Specifically, we will
assume that the $N$ states of our system are assorted into $L$
clusters, with ``strong'' links within a given cluster, $R(\sigma |
\sigma^\prime) \sim 1$, and ``weak'' links between states belonging
to different clusters, $R(\sigma | \sigma^\prime) \sim \varepsilon
\ll 1$, as illustrated in figure~\ref{bforecoarse} for $N=9$, $L=3$.
We further assume that each separate cluster is (internally) connected,
and also that the entire system is connected.
Finally, we assume that there is at
most one microscopic transition connecting any pair of states in the
system\footnote{The discussion can be generalized to several
possible transitions easily. We use this assumption since we find
the idea of a microscopic level, where the dynamics is simpler,
appealing.}. (It will become clear that coarse-graining
does not preserve this property.)
A system satisfying these conditions is characterized by two
widely separated time scales,
corresponding to the fast dynamics of relaxation within each cluster
($\tau_f \sim 1$),
and the slow redistribution of probability among the clusters
($\tau_s \sim \varepsilon^{-1}$).

The microscopic evolution of our system is fully described
by a sequence of transitions from state to state,
$\sigma\rightarrow\sigma^\prime\rightarrow\sigma^{\prime\prime}\cdots$,
along with the times at which these transitions occur.
We introduce coarse graining by imagining that we monitor
this evolution using a low-resolution apparatus, capable of distinguishing
different clusters, but not different states within a cluster.
The dynamics we observe then consists of transitions from cluster to
cluster, $c\rightarrow c^\prime\rightarrow c^{\prime\prime}\cdots$,
which provides only partial information regarding the underlying
microscopic trajectory. In general this low-resolution dynamics is
not a Markov process, and cannot be described by a master equation
of the form given by equation~(\ref{eqmaster}). We aim to
investigate whether this coarse-grained sequence satisfies a
meaningful fluctuation theorem.

\begin{figure}[t]
\center{\includegraphics[scale=0.7,width=12cm]{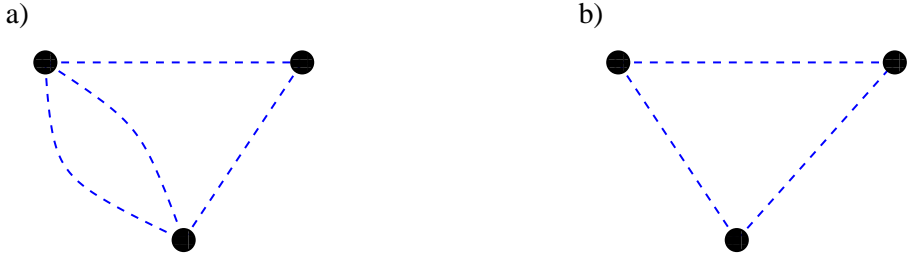}}
\caption{\label{afterc} The structure of possible simple systems,
obtained by coarse-graining the system depicted in figure
\protect{\ref{bforecoarse}}.}
\end{figure}

Let us imagine, for a moment, that such a coarse-graining
transformation maps the original system onto a smaller system, whose
states correspond to the clusters of the original system. For
instance, the system depicted in figure~\ref{bforecoarse} is mapped
onto a 3-state system. Note that there may be several different
transitions between a pair of clusters. The ability of the
low-resolution apparatus to distinguish between these transitions
determines the structure of the simpler system. For instance, the
system depicted in figure~\ref{bforecoarse}  will be mapped to the
one depicted in figure~\ref{afterc}a when the apparatus can
distinguish between the transitions connecting the same clusters.
Alternatively, when the apparatus can only distinguish the clusters,
but can not be used to identify the transitions, the system will be
mapped to the one depicted in figure~\ref{afterc}b.

In order to give a more quantitative meaning to the notion of
coarse-graining, we now develop a perturbation theory that maps our
original $N$-state stochastic jump process onto a {\it reduced}
$L$-state stochastic jump process; each state of the reduced system
represents one of the clusters in the original system. Such a
mapping is useful only if it preserves relevant dynamical
properties. In our case this means the reduced dynamics ought to
provide a good approximation of the low-resolution dynamics of the
original system. In what follows we show that the structure that we
have assumed above leads to a separation between fast and slow decay
modes of equation~(\ref{eqmaster}). The latter are then used to set
the transition rates of the reduced, $L$-state process, in such a
way that this process does, indeed, faithfully reproduce the
low-resolution dynamics of the original system, to leading order in
$\varepsilon$.

For systems with the cluster structure depicted in figure
\ref{bforecoarse}, we decompose the transition matrix, ${\bf R}$:
\begin{equation}
\label{eq:R.R0.eR1}
 R (\sigma | \sigma^\prime) = R^{(0)} (\sigma | \sigma^\prime) +\varepsilon  R^{(1)}
 (\sigma | \sigma^\prime).
\end{equation}
Here ${\bf R}^{(0)}$ is block-diagonal, with each block
corresponding to one cluster [see equation~(\ref{eq:matrixR0}) for
an example]. The non-diagonal elements of $\varepsilon {\bf
R}^{(1)}$ are the weak links connecting states belonging to
different clusters, while the diagonal elements of $\varepsilon {\bf
R}^{(1)}$ are adjusted so that equation (\ref{condition}) holds
separately for ${\bf R}^{(0)}$ and ${\bf R}^{(1)}$. We assume that
the dependence of ${\bf R}$ on $\varepsilon$ resides entirely in the
factor that multiplies ${\bf R}^{(1)}$ in the above decomposition.
Thus all non-vanishing elements of ${\bf R}^{(0)}$ and ${\bf
R}^{(1)}$ are assumed to be order unity. We then treat $\varepsilon
{\bf R}^{(1)}$ as a small perturbation to ${\bf R}^{(0)}$.

To keep considerations simple, let us assume that both ${\bf R}$
and ${\bf R}^{(0)}$ have full sets of left and right eigenvectors,
satisfying bi-orthogonality.
Thus for ${\bf R}$ we have
\begin{eqnarray}
\label{fulleigen}
{\bf R} {\bf v}_n & = & \Lambda_n {\bf v}_n \nonumber \\
{\bf w}_n^T {\bf R} & = & \Lambda_n {\bf w}_n^T \nonumber \\
{\bf w}^T_m \cdot {\bf v}_n &=& \delta_{mn},
\end{eqnarray}
with $m,n=1,2,\cdots,N$.
The same relations hold among the eigenvalues and the left and right
eigenvectors of ${\bf R}^{(0)}$, denoted by
$\Lambda_n^{(0)}$,  ${\bf w}_n^{(0)}$ and ${\bf v}_n^{(0)}$.

The transition matrix ${\bf R}$ has one vanishing eigenvalue, $\Lambda_1 = 0$.
All other eigenvalues are negative.
The corresponding right eigenvector ${\bf v}_1$ is proportional to the steady state
distribution, and the left eigenvector ${\bf w}_1$ expresses conservation of probability.
Since bi-orthogonality does not fix the normalization of these eigenvectors,
we choose, for convenience, ${\bf v}_1={\bf P}^{\rm s}$ and
${\bf w}_1^T = (1,1,\cdots,1)$.

To carry out our perturbation analysis,
we expand eigenvectors and eigenvalues in powers
of $\varepsilon$,
\begin{equation}
{\bf v}_n = \sum_{l=0}^\infty \varepsilon^l {\bf
v}_n^{(l)}
\quad,\quad
{\bf w}_n = \sum_{l=0}^\infty \varepsilon^l {\bf
w}_n^{(l)}
\quad,\quad
\Lambda_n = \sum_{l=0}^\infty \varepsilon^l
\Lambda_n^{(l)} \quad ,
\end{equation}
and we match powers of $\varepsilon$ in equation~(\ref{fulleigen}).
At leading order we get
\begin{equation}
{\bf R}^{(0)} {\bf v}_n^{(0)} = \Lambda_n^{(0)} {\bf v}_n^{(0)},
\end{equation}
and similarly for ${\bf w}_n^{(0)}$.
Making use of the block-diagonal structure of ${\bf R}^{(0)}$,
and diagonalizing each block separately,
we obtain eigenvectors ${\bf v}_n^{(0)}$ and ${\bf w}_n^{(0)}$
that inherit this structure:
each eigenvector has non-vanishing elements only
in one block, corresponding to one cluster.

The block-diagonal structure of ${\bf R}^{(0)}$ describes $L$
mutually isolated clusters, each of which supports its own steady
state and corresponding vanishing eigenvalue. Thus ${\bf R}^{(0)}$
has $L$ vanishing eigenvalues, $\Lambda_n^{(0)} =0$ for $n=1,\cdots
,L$. The corresponding right eigenvectors, ${\bf v}_n^{(0)}$, are
composed of blocks of vanishing elements with the steady state of
the $n$'th cluster, ${\mathfrak P}^{\rm s} (n; \sigma)$ inserted at
the appropriate place. (Here $\sigma$ runs over states belonging to
the $n$'th cluster.) The corresponding left eigenvectors, ${\bf
w}_n^{(0)}$, have a similar structure, with the elements ${\mathfrak
P}^{\rm s} (n; \sigma)$ replaced by $1$'s. The remaining $N-L$
eigenvectors of ${\bf R}^{(0)}$ have negative eigenvalues of order
unity.


In first-order perturbation theory, the term $\varepsilon {\bf
R}^{(1)}$ in equation~(\ref{eq:R.R0.eR1}) shifts each eigenvalue by
an amount of order $\varepsilon$. While this represents a relatively
small change in the non-vanishing eigenvalues
$\Lambda_{L+1}^{(0)},\cdots,\Lambda_N^{(0)}$, it crucially lifts the
degeneracy among the first $L$ (vanishing) eigenvalues. These are
replaced by a single vanishing eigenvalue (corresponding to the
steady state of the fully connected system) and $L-1$ negative
eigenvalues of order $\varepsilon$, which describe the long time
decay of the system. Assuming for simplicity that the degeneracy is
fully lifted at first order, higher order corrections will lead only
to small further changes in the eigenvalues and eigenvectors.
Therefore, in the following we will concentrate on the application
of first-order perturbation theory to the degenerate sector, $n\le
L$.

Degenerate perturbation theory begins with the $L \times L$ matrix
\begin{equation}
\label{defr1} \Pi (n | m)  \equiv {\bf w}_n^{(0)T}  {\bf R}^{(1)}
{\bf v}_m^{(0)}
\quad,\quad
m,n=1,\cdots,L ,
\end{equation}
whose eigenvalues represent the first-order corrections, $\Lambda_n^{(1)}$,
to the vanishing eigenvalues of ${\bf R}^{(0)}$.
Because of the correspondence between
the clusters of the system and
the left/right eigenvectors in the above equation,
the quantity $\Pi (n | m)$ represents the transition rate between
clusters $m$ and $n$.
In the following we emphasize this identification by using indices
$c,c^\prime=1,\cdots,L$ to specify the clusters.

The slow decay modes of the system are determined by the matrix
$\Pi$ whose dimensionality is the number of clusters, $L$.
This matrix satisfies
\begin{equation}
\fl \sum_{c=1}^L \Pi(c | c^\prime) = \sum_{c=1}^L
\sum_{\sigma,\sigma^\prime=1}^N {w}_c^{(0)} (\sigma) {R}^{(1)}
(\sigma | \sigma^\prime){v}_{c^\prime}^{(0)} (\sigma^\prime) =
\sum_{\sigma,\sigma^\prime=1}^N R^{(1)}(\sigma | \sigma^\prime)
v_{c^\prime}^{(0)} (\sigma^\prime)=0,
\end{equation}
where we have used $\sum_{c=1}^L w_c^{(0)} (\sigma) =1$. In other
words $\Pi$ satisfies the conservation-of-probability condition
expressed (for the full system) by equation~(\ref{condition}). It is
thus meaningful to interpret the equation,
\begin{equation}
\label{coarsemaster} \frac{d \tilde{P} (c,t)}{d t} = \sum_{c^\prime}
\varepsilon \Pi (c | c^\prime) \tilde{P} (c^\prime,t),
\end{equation}
as describing a Markov jump process.
We now discuss the physical meaning that can be assigned to $\Pi$ and $\tilde{P} (c,t)$.

The separation of time scales between the $L-1$ slow and the $N-L$
fast decay modes, implies that after the decay of the latter the
probability distribution $P(\sigma,t)$ satisfies $P (\sigma ,t)
\simeq {\mathfrak P}^{\rm s} (c;\sigma) \tilde{P} (c,t)$. Thus the
probability inside each cluster is a product of its local steady
state distribution, ${\mathfrak P}^{\rm s}$, and the total
probability to find the system in the cluster, $\tilde{P} (c,t) =
\sum_{\sigma \in c}P(\sigma,t)$. Equation (\ref{coarsemaster})
approximately describes the time evolution of the latter. More
precisely, equation~(\ref{coarsemaster}) governs a {\it reduced}
($L$-state) Markov jump process, and {\it this reduced process
provides an approximate description of the transitions between
clusters within the original system}. Note that the individual
elements of ${\bf \Pi}$ differ from those of ${\bf R}^{(1)}$: for a
transition out of a cluster $c$, with the rate $\varepsilon_{+}$
appearing in $\varepsilon {\bf R}^{(1)}$, the probability current is
$\varepsilon_{+} P(\sigma,t) \simeq \varepsilon_{+} {\mathfrak
P}^{\rm s} (c;\sigma) \tilde{P} (c,t)$. This contributes a rate
$\varepsilon_{+}{ \mathfrak P}^{\rm s} (c;\sigma)$ in the
appropriate place in $\varepsilon {\bf \Pi}$.

The procedure outlined above is not the only way to define a reduced
system. An alternative approach would involve measuring the
transition rates between clusters in the steady state of the full
system, and then constructing a reduced, $L$-state Markov jump
process governed by precisely these transition rates. We expect that
this ``empirical construction'' -- based on the observed
cluster-to-cluster transition rates -- and the ``perturbative
construction'' leading to equation~(\ref{coarsemaster}), will give
reduced systems that become equivalent in the limit
$\varepsilon\rightarrow 0$.

To summarize, we have used perturbation theory to give quantitative
meaning to a coarse-graining transformation, in which a Markov jump
process among $N$ states, arranged into $L$ clusters, is replaced by
a reduced Markov jump process among $L$ states meant to represent
those clusters. In the following section, where we consider
fluctuations, it is important to bear in mind that the reduced
dynamics only approximately reproduce the statistics of transitions
between clusters of the original system. Of course, one might
improve the approximation by extending the perturbation expansion to
higher orders of $\varepsilon$; see, e.g.\ the Appendix of
reference~\cite{Geyer2006}. However, higher-order corrections to the
eigenvectors of a particular degenerate sector typically involve
{\em all} unperturbed eigenvectors, not only those within that
sector. Thus it is not generally possible to reduce the
dimensionality of the system when considering higher orders in
$\varepsilon$.

\section{Fluctuation relations for systems that can be coarse-grained}
\label{fluct}

In this section we ask whether the structure described in
section~\ref{coarse} leads to a non-trivial fluctuation relation. We
start by briefly reviewing the fluctuation relation for general
stochastic jump
processes~\cite{Lebowitz1999,Seifert2005,Imparato2007,Harris2007}.
We will then turn to systems amenable to coarse-graining.

Fluctuation relations are obtained by comparing probability
densities of microscopic paths and their time reversed counterparts. A
path, or a history, $\gamma (T)$, will denote a stochastic
trajectory, represented by an ordered sequence of transitions at given times (between
say $0$ and $T$),
\begin{equation}
\label{defsmallg} \gamma \equiv \sigma_0
\stackrel{t_1,\alpha_1}{\longrightarrow} \sigma_1
\stackrel{t_2,\alpha_2}{\longrightarrow} \sigma_2 \longrightarrow
\cdots \stackrel{t_n,\alpha_n}{\longrightarrow} \sigma_n .
\end{equation}
The $\alpha$'s in equation (\ref{defsmallg}) denote the transitions
between states. They are redundant for the original system, but
would be needed after coarse-graining, where several transitions
between clusters are possible. We have chosen to explicitly denote
the transitions in order to use similar notation for the original
and coarse-grained system. The conditional probability density (or
weight) corresponding to a path is
\begin{equation}
\label{weight} w(\gamma) = \prod_{i=0}^n \exp \left[ R (\sigma_i |
\sigma_i) (t_{i+1}-t_{i})\right] \prod_{j=1}^n R_{\alpha_j}
(\sigma_j | \sigma_{j-1}).
\end{equation}
In equation (\ref{weight}) $-R (\sigma_i | \sigma_i)$ denotes the
overall rate of probability flow out of the state $\sigma_i$, and we
set $t_0=0$ and $t_{n+1}=T$. The probability density of a path is
obtained by multiplying $w(\gamma)$ by the probability to sample its
initial conditions, $P (\sigma_0 , t=0)$.

To each path we can define a time reversed counterpart
$\bar{\gamma}(t) \equiv \gamma (T-t)$, starting at $\sigma_n$ at
$t=0$, and ending at $\sigma_0$ at $t=T$. Since the path $\gamma$
and its counterpart $\bar{\gamma}$ reside in the same states for the
same lengths of time their exponential survival probability factors
in equation (\ref{weight}) are identical. As a result
\begin{equation}
\label{weightratio} \frac{{\cal P}(\gamma)}{{\cal P}(\bar{\gamma})}
= \frac{P(\sigma_0 , t=0) w(\gamma)}{P(\sigma_n , t=0)
w(\bar{\gamma})} =\frac{P (\sigma_0 , t=0) }{P (\sigma_n , t=0) }
\prod_{i=1}^n \frac{R_{\alpha_i} (\sigma_i |
\sigma_{i-1})}{R_{\alpha_i} (\sigma_{i-1} | \sigma_{i})}.
\end{equation}
Taking the logarithm of both sides yields
\begin{equation}
\ln \frac{{\cal P}(\gamma)}{{\cal P}(\bar{\gamma})}
=
\ln \frac{P (\sigma_0 , t=0) }{P (\sigma_n , t=0) }
+
\sum_{i=1}^n
\ln \frac{R_{\alpha_i} (\sigma_i |
\sigma_{i-1})}{R_{\alpha_i} (\sigma_{i-1} | \sigma_{i})} \equiv B + \sum_{i=1}^n \delta S_i .
\end{equation}
The right side includes a boundary term, $B$, arising from the initial and final
states,
and a sum of contributions from the $n$ transitions.
If we interpret
\begin{equation}
\label{eq:deltaS}
\delta S =
\ln \frac{ R_{\alpha} (\sigma | \sigma^\prime) }
{ R_{\alpha} (\sigma^\prime  | \sigma) }
\end{equation}
as the amount of entropy that is produced
when the system makes a transition $\alpha$ from $\sigma^\prime$ to $\sigma$~\cite{Lebowitz1999,Seifert2005,Harris2007},
then
\begin{equation}
\label{eq:DS}
\Delta S = \sum_{i=1}^n \delta S_i
\end{equation}
is the net entropy production associated with the trajectory $\gamma$,
and we have
\begin{equation}
\label{eq:preFT}
\ln \frac{{\cal P}(\gamma)}{{\cal P}(\bar{\gamma})}
=
B + \Delta S.
\end{equation}


Identities of the form of equation~(\ref{eq:preFT}) are at the heart
of both {\it steady-state} and {\it transient} fluctuation
relations~\cite{Harris2007}. The distinction between the two cases,
roughly, amounts to whether one chooses to consider the limit
$T\rightarrow\infty$ (in which case the relative contribution of $B$
can be neglected, at least for finite systems~\cite{Lebowitz1999}),
or rather to argue that the boundary term $B$ represents the net
change in the entropy of the system (in which case the sum on the
right represents the entropy produced in the system and its
surroundings~\cite{Seifert2005}). In either situation the desired
fluctuation relation follows in a more or less straightforward
manner.

For our purposes the distinction between the transient and steady
state cases is not of central importance. Rather, we aim to
investigate whether a relationship similar to
equation~(\ref{eq:preFT}) can be obtained for systems of the type
studied in section~\ref{coarse}. To clarify this issue, let us
return to the situation introduced briefly in section~\ref{coarse},
in which we monitor the evolution of our system using a
low-resolution apparatus that distinguishes between clusters, but
not between states within a cluster. Under these circumstances we do
not observe the full trajectory, $\gamma$, but rather a
coarse-grained version $\Gamma$ (defined below) that describes only
transitions between clusters. Can we expect this trajectory to
satisfy a fluctuation relation? In what follows we use the
perturbation analysis of section~\ref{coarse} to define a
coarse-grained entropy production, $\Delta S^{\rm CG}$, that differs
from $\Delta S$ above, and is expressed in terms of the {\it
observed} transition rates. We argue that this quantity
approximately satisfies a fluctuation relation, and that the
approximation improves as the separation of time scales becomes more
pronounced (i.e.\ as $\varepsilon\rightarrow 0$). The deviations
from this (approximate) relation are more pronounced for extreme
values of $\Delta S^{\rm CG}$, which select rare trajectories that,
in some sense, do not respect the separation of time scales. These
deviations reflect the non-Markovian nature of the observed,
coarse-grained dynamics, and can be viewed as evidence of the
existence of unobserved degrees of freedom.

The reduced, $L$-state system defined in section~\ref{coarse} is a
stochastic, Markov jump process that satisfies an exact fluctuation
relation. The entropy produced during a transition $\alpha$ from $c$
to $c^\prime$ is given in this case by
\begin{equation}
\label{eq:dS_red}
\delta S^{\rm red} =
\ln \frac{\Pi_{\alpha} ( c | c^\prime )}{\Pi_{\alpha}( c^\prime | c )} ,
\end{equation}
and the net entropy production associated with a trajectory is just
a sum of such contributions. To simplify notations, we
assume here, and in what follows, the scenario where different
transitions between the same clusters are distinguishable. The
following considerations can be easily modified to apply to
coarse-grained dynamics with indistinguishable transitions.

Let us now return to the full, $N$-state system,
but assume the conditions of a low-resolution apparatus.
Suppose the evolution of the system from $t=0$ to $t=T$ is described by
a trajectory $\gamma$.
To this trajectory let us assign a {\it coarse-grained} entropy production
$\Delta S^{\rm CG}[\gamma]$, which is a sum of contributions from jumps between
different clusters.
Specifically, when the system makes a transition $\alpha$ from cluster $c^\prime$
to cluster $c$, $\Delta S^{\rm CG}$ is updated by an amount
\begin{equation}
\label{coarseentropy}
\delta S^{\rm CG} =
\ln \frac{\Pi_{\alpha} ( c | c^\prime )}{\Pi_{\alpha}( c^\prime | c )} .
\end{equation}
There are no changes to $\Delta S^{\rm CG}$ arising from
(unobserved) jumps within clusters. Thus
equation~(\ref{coarseentropy}) defines a coarse-grained entropy
production for the $N$-state system, under low resolution, by
borrowing the definition that arises in the reduced, $L$-state
Markov system, equation~(\ref{eq:dS_red}). We now argue that
\begin{equation}
\label{eq:DSCG} \Delta S^{\rm CG} = {\sum_k}^\prime \delta S_k^{\rm
CG}
\end{equation}
(the sum includes only cluster-to-cluster transitions) satisfies an approximate
fluctuation relation.

Since only the transitions between different clusters are observed under
low-resolution conditions, let us define a coarse-grained path
\begin{equation}
\label{defcoarsepath} \Gamma \equiv \left\{ \gamma | c_0
\stackrel{\tau_1,\alpha_1}{\longrightarrow} c_1
\stackrel{\tau_2,\alpha_2}{\longrightarrow} \cdots
c_{n^\prime}\right\}.
\end{equation}
We interpret $\Gamma$ as the collection of all paths $\gamma$, of
duration $T$, that start in the cluster $c_0$ at time $0$, make the
transition $\alpha_1$ to cluster $c_1$ at time $\tau_1$ and so
forth. Thus while the inter-cluster transitions of
all trajectories in $\Gamma$ are identical,
the intra-cluster transitions typically differ among these trajectories.
Note that we retain information on the identity ($\alpha$)
of a transition between two clusters, if only to identify the microscopic states involved.
Moreover, since $\Delta S^{\rm CG}$ has the same value for every
$\gamma\in\Gamma$,
we take this value to define the coarse-grained entropy production
associated with $\Gamma$.

The probability density assigned to the coarse-grained path
$\Gamma$ is
\begin{equation}
{\cal P}(\Gamma) = \sum_{\gamma \in \Gamma} {\cal P}(\gamma).
\end{equation}
This is computed by summing over all possible
combinations of transitions inside the cluster and integrating over
their corresponding transition times, keeping the inter-cluster
transitions fixed. Each coarse-grained trajectory has a time
reversed counterpart $\bar{\Gamma}$, which passes
through the clusters in the reversed order, at the appropriate times.
Let us now compare the statistical weights associated
with such a {\it conjugate pair} of coarse-grained trajectories, $\Gamma$
and $\bar\Gamma$.

The weight of a coarse-grained trajectory is a product of factors
(partial weights) resulting from the different segments of the
microscopic trajectories. A {\it segment} here refers to the
interval of time -- bracketed by a pair of transition times, say
$\tau_k$ and $\tau_{k+1}$ -- during which the system remains in
exactly one cluster. [E.g.\ equation~(\ref{defcoarsepath}) specifies
a path with $n^\prime+1$ segments.]

\begin{figure}[t]
\center{\includegraphics[scale=0.5]{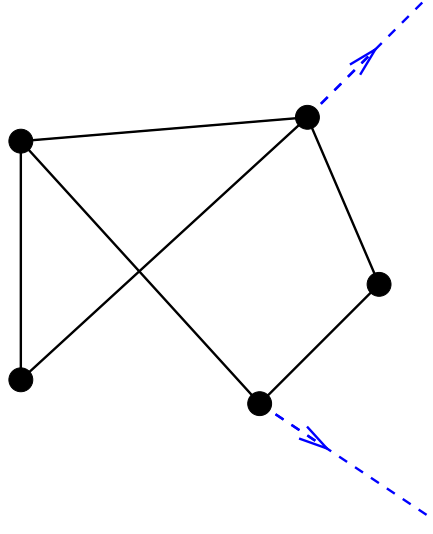}}
\caption{\label{defmu} A cluster of states with some additional
outgoing transitions. Stochastic motion in such a system is related
to partial contributions to weights of coarse-grained trajectories,
see text.}
\end{figure}

To compute the partial weight of a given segment, consider the
system shown in figure~\ref{defmu},
with five states and two outgoing
links. The latter represent the decay of the probability
to remain within this system: when a trajectory makes a transition
along such a link, it departs the system and does not return.  Let
$\mu (\sigma,t | \sigma^\prime,t_0)$
denote the conditional probability to find the system at
$(\sigma,t)$, given that it was at $(\sigma^\prime,t_0)$.

The partial weights discussed above are contributions from
trajectory segments that perform exactly this type of stochastic
dynamics. By specifying $\Gamma$ one specifies all the transitions
between the clusters. Given such a transition, $\alpha_k$, let us
denote the microscopic states before and after the transition by
$\sigma_k^-$ and $\sigma_k^+$, respectively. The contribution from
all trajectory segments entering $c_k$ at $(\sigma_k^+,\tau_k)$ and
leaving this cluster for the first time at
$(\sigma_{k+1}^-,\tau_{k+1})$, is just
$\mu(\sigma_{k+1}^-,\tau_{k+1}| \sigma_k^+,\tau_k)$. \footnote{Note
that each cluster has its own conditional probability distribution
$\mu$. We will not introduce new notation to explicitly
differentiate between the different functions $\mu$, since the
identity of the cluster is already specified in the microscopic
states that appear as arguments of $\mu$.}

For the systems studied here, the separation of time scales suggests
that the system is likely to relax rapidly within a cluster before departing
that cluster, which in turn suggests a simple approximation:
\begin{equation}
\label{approxmu} 
\mu (\sigma_{k+1}^-,\tau_{k+1} | \sigma_k^+, \tau_k) \simeq \exp
\left[ \varepsilon \Pi (c_k,c_k) (\tau_{k+1} -\tau_k)\right]
{\mathfrak P}^{\rm s} (c_k;\sigma_{k+1}^-) .
\end{equation}
Here $-\varepsilon \Pi (c_k, c_k)$ is the approximate rate of escape
from $c_k$, assuming the system has relaxed within this cluster,
while ${\mathfrak P}^{\rm s} (c_k;\sigma_{k+1}^-)$ is the steady
state probability to be at $\sigma_{k+1}^-$ in the isolated cluster
$c_k$. At this level of approximation, the weight $\mu$ does not
depend on the initial state $\sigma_k^+$. We defer discussion of
possible corrections to equation (\ref{approxmu}) to a later stage.
The probability density of a coarse-grained trajectory is then
\begin{eqnarray}
\label{coarseweight} \fl {\cal P}(\Gamma)  &  =  &
\left[\sum_{\sigma_0 \in c_0} P (\sigma_0 , t=0) \mu (\sigma_1^-,
\tau_1 | \sigma_0,0)\right] \prod_{k=1}^{n^\prime-1} \mu
(\sigma_{k+1}^-,\tau_{k+1} | \sigma_k^+,\tau_k)
\\ \fl &   \times & \left[ \sum_{\sigma \in c_{n^\prime}} \mu (\sigma,T |
\sigma_{n^\prime}^+,\tau_{n^\prime})\right] \prod_{k=1}^{n^\prime}
\varepsilon R^{(1)}_{\alpha_k} (\sigma_k^+ | \sigma_k^-) \nonumber.
\end{eqnarray}
The two factors in square brackets, corresponding to the initial and
final clusters, account for the fact that the initial and final
microstates are not fully specified by $\Gamma$. Substituting
equation (\ref{approxmu}) into equation (\ref{coarseweight}) leads
to
\begin{eqnarray}
\label{eq:approxMarkov} \fl {\cal P}(\Gamma) & \simeq & \tilde{P}
(c_0 , t=0) \prod_{k=0}^{n^\prime} \exp \left[ \varepsilon \Pi (c_k
| c_k) (\tau_{k+1}-\tau_k)\right] \prod_{k=1}^{n^\prime} \left[
\varepsilon R^{(1)}_{\alpha_k} (\sigma_k^+| \sigma_k^-) {\mathfrak
P}^{\rm s} (c_k;\sigma_k^-)\right] \nonumber \\
\fl & = & \tilde{P} (c_0 , t=0) \prod_{k=0}^{{n^\prime}} \exp \left[
\varepsilon \Pi (c_k | c_k) (\tau_{k+1}-\tau_k)\right]
\prod_{k=1}^{n^\prime} \varepsilon \Pi_{\alpha_k} (c_{k}| c_{k-1}),
\end{eqnarray}
where $\Pi_{\alpha_k} (c_{k} | c_{k-1})$ is the contribution of the
process $\alpha_k$, to the element $(c_{k} , c_{k-1})$ of the
reduced transition matrix $\Pi$ (see section~\ref{coarse}). The
right side of this equation is exactly the weight of a trajectory in
the reduced, Markovian system whose time evolution obeys equation
(\ref{coarsemaster}). In other words, equation
(\ref{eq:approxMarkov}) states that the coarse-grained trajectories
$\Gamma$ are described by approximately the same statistics as those
of the reduced, $L$-state Markovian system. Evaluating the ratio of
weights for a conjugate pair $\Gamma$ and $\bar\Gamma$, and applying
our definition of $\Delta S^{\rm CG}$ [equation~(\ref{eq:DSCG})], we
get
\begin{equation}
\label{ratioofcoarse} \frac{{\cal P}(\Gamma)}{{\cal
P}(\bar{\Gamma})} \simeq
\frac{\tilde{P} (c_0 ,t=0)}{\tilde{P} (c_{n^\prime},t=0)} \cdot \exp
\left( \Delta S^{\rm CG}[\Gamma] \right).
\end{equation}
(The factors containing the diagonal elements of $\Pi$, appearing in
equation~(\ref{eq:approxMarkov}), cancel in this ratio.) At this
point we have arrived at an analogue of equation~(\ref{eq:preFT}),
the crucial relation from which both transient and steady-state
fluctuation relations are obtained. Our analysis thus reveals that
the coarse-grained entropy production $\Delta S^{\rm CG}$ -- which
is defined for the full, $N$-state system under conditions of low
resolution, but which was motivated by a comparison with the
reduced, $L$-state system of section~\ref{coarse} -- satisfies an
approximate fluctuation relation. We will rederive this result,
using different considerations, in the next section.

Let us briefly take a closer look at the coarse-grained entropy,
equation~(\ref{coarseentropy}). Comparing the definitions of $\delta
S^{\rm CG}$ and $\delta S$, we find
\begin{equation}
\label{interpretentropy} \delta S^{\rm CG}(\alpha) = \delta S
(\alpha) + \ln \frac{{\mathfrak P}^{\rm s} (c^\prime ;
\sigma^-)}{{\mathfrak P}^{\rm s}(c; \sigma^+)}.
\end{equation}
Entropy changes in stochastic jump processes are often separated
into two contributions with different physical
interpretations~\cite{Zia2006}. The first term on the right hand
side of equation (\ref{interpretentropy}) expresses the change of
the entropy of a thermal medium that is in contact with the
system~\cite{Lebowitz1999,Gaspard2004,Imparato2007,
Harris2007,Zia2006,Schmiedl2006,Schnakenberg1976}. The second term
is related to changes in the entropy of the system itself, if we
interpret $-\ln P(\sigma(t),t)$ as the entropy of the system at time
$t$ along a stochastic trajectory $\sigma(t)$~\cite{Seifert2005}.
Equation (\ref{interpretentropy}) indicates that coarse-graining
does not respect the distinction between the microscopic medium and
system entropies. Both contribute to changes in the coarse-grained
entropy.

Note also that transitions within the clusters contribute to $\Delta
S$ [equation~(\ref{eq:DS})], but not to $\Delta S^{\rm CG}$
[equation~(\ref{eq:DSCG})]. In fact, there may be situations where
the full, $N$-state system relaxes to a steady state characterized
by non-zero stationary currents within the clusters, but transitions
between clusters satisfy detailed balance. This would correspond to
a state with a non-vanishing rate of entropy production $\Delta S$,
but a vanishing rate of coarse-grained entropy production, $\Delta
S^{\rm CG}$.

The claims made in previous paragraphs rely on
equation~(\ref{approxmu}), which is an approximation. We should
therefore consider possible corrections to equation (\ref{approxmu})
and their effects on the fluctuation relation. We will use heuristic
rather than rigorous arguments to estimate these corrections.


The corrections to equation (\ref{approxmu}) depend on the time
difference $\delta \tau_k \equiv \tau_{k+1}-\tau_k$. For typical
time differences, comparable to the lifetime in the cluster,
equation (\ref{approxmu}) is a good approximation, but we can expect
it to have errors of order $\varepsilon$. For very short ($\delta
\tau_k \sim 1$) and very long ($\delta \tau_k \sim ln \varepsilon /
\varepsilon$) time differences we may expect large errors. However,
the probability of such time differences is itself of order
$\varepsilon$. Such events are not typical.

The (coarse-grained) steady state fluctuation relation applies to
the long time behavior of $(\varepsilon T)^{-1} \ln \left[ P(\Delta
S^{\rm CG} = q \varepsilon T)/ P(\Delta S^{\rm CG} = -q \varepsilon
T )\right]$. The only source of errors will then appear in sums over
terms of the form $\ln
\mu(\sigma,\tau|\sigma^\prime,\tau^\prime)/\mu(\sigma^\prime,T-\tau^\prime|\sigma,T-\tau)$.
The number of terms in the sum is proportional to the number of
transitions between clusters, which in turn scales as $\varepsilon
T$. Let us consider first the corrections for a typical trajectory.
The corrections are a combination of typical terms with small errors
and rare terms with a large error, as discussed in the previous
paragraph. The resulting errors in estimating the entropy production
is small, and scales as $\varepsilon^2 T$.

The steady state fluctuation relation does not deal only with
typical trajectories. In fact, one examines trajectories whose
entropy production rate is (approximately) fixed. This will pick out
trajectories which may be very unlikely.
We {\em assume} that for a finite range of entropy production rates
the relevant trajectories are such that the time differences for
which the approximation made in equation (\ref{approxmu}) break down
are still unlikely (with probability which scales as $\varepsilon$).
With this assumption, the coarse grained-entropy satisfies
\begin{equation}
\label{steadystatefr} \lim_{T \rightarrow \infty}
\frac{1}{\varepsilon T} \ln \frac{P(\Delta S^{\rm CG} =
\varepsilon q T)}{P(\Delta S^{\rm CG} =- \varepsilon q T)} = q +
O(\varepsilon),
\end{equation}
for some finite range of $q$ values\footnote{The numerical results
presented in section \ref{example} hint that the error term in
equation (\protect\ref{steadystatefr}) depends on $q$ in a
non-trivial manner. }. Note that one can not expect this relation to
hold for any coarse-grained entropy production rate $q$. Taking this
rate to be "large enough" will eventually pick trajectories whose
typical time differences are of order $\tau_f$. These trajectories
will never spend enough time in clusters to equilibrate.

\section{Illustrative Example}
\label{example}

In this section we use a simple example to illustrate fluctuation
relations in a coarse-grained setting, and to gain some intuition
for the admittedly abstract considerations of the previous section.
We want to use the
simplest system possible whose reduced counterpart can
exhibit out-of-equilibrium steady states. This implies at
least three clusters of states.

Consider a system with eight states, $N=8$, organized
into three clusters, $L=3$. The transition matrix describing the
full system is
\begin{equation}
{\bf R} = {\bf R}^{(0)}+\varepsilon {\bf R}^{(1)},
\end{equation}
with
\begin{equation}
\label{eq:matrixR0}
{\bf R}^{(0)} = \left( \begin{array}{cccccccc} {\bf -3} & {\bf 3} &
{\bf 2} & 0 & 0 & 0 & 0 & 0 \\ {\bf 2} & {\bf -8} & {\bf 3} & 0 & 0 & 0 & 0 &0 \\
{\bf 1} & {\bf 5} & {\bf -5} & 0 & 0 & 0 & 0 & 0 \\ 0 & 0 & 0 & {\bf -2} & {\bf 3}& 0 & 0 & 0 \\
0 & 0 & 0 & {\bf 2} & {\bf -3} & 0 & 0 & 0 \\ 0 & 0 & 0 & 0 & 0 & {\bf -4} & {\bf 2} & {\bf 2} \\
0 & 0 & 0 & 0 & 0 & {\bf 3} & {\bf -3} & {\bf 6} \\ 0 & 0 & 0 & 0 &
0 & {\bf 1} & {\bf 1} & {\bf -8}
\end{array} \right),
\end{equation}
and
\begin{equation}
{\bf R}^{(1)} = \left( \begin{array}{cccccccc} 0 & 0 & 0 & 0 & 0 & 0
& 0 & 0 \\ 0 & {\bf -4} & 0 & {\bf 2} & 0 & 0 & 0 & 0 \\ 0 & 0 &
{\bf -1} & 0 & 0 & 0 & {\bf 1} & 0 \\ 0 & {\bf 4} & 0 & {\bf -2} & 0 & 0 & 0 & 0 \\
0 & 0 & 0 & 0 & {\bf -5} & {\bf 3} & 0 & 0  \\ 0 & 0 & 0 & 0 & {\bf 5} & {\bf -3} & 0 & 0 \\
0 & 0 & {\bf 1} & 0 & 0 & 0 & {\bf -1} & 0 \\ 0 & 0 & 0 & 0 & 0 & 0
& 0 & 0
\end{array} \right).
\end{equation}
 (The non-vanishing rates are in bold-face.)
 Numbering the states of the full system from 1 to 8,
 the three clusters are: $(1,2,3)$, $(4,5)$, and $(6,7,8)$.
 We see that ${\bf R}$ has
 the structure described in section~\ref{coarse}:
 all off-diagonal elements of ${\bf R}^{(0)}$ are between states within a given
 cluster, and
 all off-diagonal elements of ${\bf R}^{(1)}$ are between states belonging
 to different clusters.
We begin our analysis of this system by constructing the reduced,
three-state Markov system.

The steady state distributions for the (unconnected) clusters
appearing in ${\bf R}^{(0)}$ are given by the null vectors of the
three matrices along the block diagonal: ${\mathfrak P}^{\rm s} (1;
\sigma \in 1) = \frac{1}{56} ( 25,13,18)$, ${\mathfrak P}^{\rm s}(2;
\sigma \in 2) = \frac{1}{5} (3,2)$ and ${\mathfrak P}^{\rm s}(3;
\sigma \in 3)=\frac{1}{9} (3,5,1)$. From these we easily construct
the eigenvectors spanning the null space of ${\bf R}^{(0)}$. Using
equation (\ref{defr1}) we then obtain
\begin{equation}
\label{cgsystem} {\bf \Pi} = \left( \begin{array}{ccc} - \frac{5}{4}
&
\frac{6}{5} & \frac{5}{9} \\ \frac{13}{14} & - \frac{16}{5} & 1 \\
\frac{9}{28} & 2 & - \frac{14}{9} \end{array} \right).
\end{equation}
This transition matrix defines the dynamics of the three-state reduced system,
and approximately describes the cluster-to-cluster evolution of the full, eight-state system.

To evaluate the probability distribution of
entropy production rates in the limit of large observation time, $T$,
direct simulations are of limited use, as
the asymptotic probability to observe a
specific {\it average} entropy production rate decays exponentially with $T$.
Thus obtaining good statistics becomes problematic.
Following Lebowitz and Spohn~\cite{Lebowitz1999}
(see also \cite{Imparato2007}), we consider the large deviation
function
\begin{equation}
f (q)\equiv \lim_{T \rightarrow \infty} \frac{1}{ \varepsilon T} \ln
P (\Delta S^{\rm CG}=q \varepsilon T , T),
\end{equation}
and its Legendre transform $g (\lambda)$. Here $q$ denotes the
observed entropy production rate, while $P(\Delta S^{\rm CG})$ is
the probability distribution of this entropy production. To
differentiate between the large deviation function of the reduced
and coarse-grained system we will denote the former (latter) by a
superscript $red$ ($CG$).

Lebowitz and Spohn have shown that $g(\lambda)$ is the largest
eigenvalue of an operator ${\cal H} (\lambda)$ which governs the
time evolution of averages of the type $\left< e^{- \lambda \Delta
S} \right>$. The diagonal elements of ${\cal H}$ are those of ${\bf
R}$, while non-diagonal elements are multiplied by a factor related
to $e^{-\lambda \Delta S}$ where the entropy change is that of the
transition corresponding to this non-diagonal element. We refer for
references \cite{Lebowitz1999} and \cite{Imparato2007} for the
derivation.

For the reduced system, whose transition matrix is given by equation
(\ref{cgsystem}), this operator is given by
\begin{equation}
\label{calhtilde} \tilde{\cal H} (\lambda) = \varepsilon \left(
\begin{array}{ccc} - \frac{5}{4} & \frac{6}{5}
\left(\frac{65}{84}\right)^\lambda & \frac{5}{9}
\left(\frac{81}{140}\right)^\lambda \\ \frac{13}{14}
\left(\frac{84}{65}\right)^\lambda & - \frac{16}{5} & 2^\lambda \\
\frac{9}{28} (\frac{140}{81})^\lambda & 2^{1-\lambda} & -
\frac{14}{9}
\end{array} \right).
\end{equation}
This operator has the property $\tilde{\cal H} (\lambda)=\tilde{\cal
H}^T (1-\lambda)$, which implies the steady state
(or Gallavotti-Cohen) fluctuation relation $f^{\rm red}(q)-f^{\rm
red}(-q)=q$~\cite{Gallavotti1995,Lebowitz1999}. Note that the
dependence of $\tilde{\cal H}$ on $\varepsilon$ is trivial. To
cancel this dependence we define $g^{\rm red}(\lambda)$ to be the
maximal eigenvalue of $\tilde{\cal H}$ divided by $\varepsilon$.

Let us turn back to the original system. We have defined a
coarse-grained entropy which changes only during stochastic
transitions connecting different clusters, see equation
(\ref{coarseentropy}). One can apply the formalism developed by
Lebowitz and Spohn to study the distribution function of entropy
production for the coarse-grained dynamics. The derivation follows
closely that of \cite{Imparato2007}. Again, the Legendre transform
of the large deviation function is given by the largest eigenvalue
of an operator (divided by $\varepsilon$). The relevant operator is
\begin{equation} \fl
\label{calh} {\cal H} (\lambda) = \left( \begin{array}{cccccccc}
{\bf -3} & {\bf 3} & {\bf 2} & 0 & 0 & 0 & 0 & 0 \\ {\bf 2} & {\bf
-8 -4 \boldsymbol{\varepsilon } }& {\bf 3} &
{\bf 2 \boldsymbol{ \varepsilon } \left( \frac{65}{84}\right)^{\boldsymbol{\lambda }}} & 0 & 0 & 0 & 0 \\
{\bf 1} & {\bf 5} & {\bf -5 -\boldsymbol{\varepsilon } } & 0 & 0 & 0
& {\bf \boldsymbol{ \varepsilon } \left( \frac{81}{140}\right)^{\boldsymbol{\lambda }}} & 0 \\
0 & {\bf 4 \boldsymbol{\varepsilon} \left(
\frac{84}{65}\right)^{\boldsymbol{\lambda}}} & 0 & {\bf -2 - 2
\boldsymbol{\varepsilon}} & {\bf 3} & 0 & 0 & 0
\\ 0 & 0 & 0 & {\bf 2} & {\bf -3 - 5 \boldsymbol{\varepsilon}} & {\bf 3 \boldsymbol{\varepsilon} 2^{\boldsymbol{\lambda }}} & 0
& 0 \\ 0 & 0 & 0 & 0 & {\bf 5  \boldsymbol{\varepsilon}
2^{-\boldsymbol{\lambda}}} & {\bf -4 -3 \boldsymbol{\varepsilon}} &
{\bf 2} & {\bf 2}
\\ 0 & 0 & {\bf \boldsymbol{\varepsilon} \left(\frac{140}{81} \right)^{\boldsymbol{\lambda }}} & 0 & 0
& {\bf 3} & {\bf -3- \boldsymbol{\varepsilon} }& {\bf 6} \\ 0 & 0 &
0 & 0 & 0 & {\bf 1} & {\bf 1} & {\bf -8}
\end{array} \right).
\end{equation}
Note that only elements which connect different clusters depend on
$\lambda$. This results from the fact that the coarse-grained
entropy changes only during these transitions. It is important to
notice that ${\cal H} (\lambda) \neq {\cal H}^T (1-\lambda)$. This
means that the coarse-grained entropy does not satisfy an {\em
exact} fluctuation relation. It does, however satisfy an approximate
one.

Consider ${\cal H}$ for small values of $\varepsilon$. When
$\varepsilon=0$ we find that ${\cal H} = {\bf R}^{(0)}$. We have
already studied the eigenvalues and eigenvectors of ${\bf R}^{(0)}$
in section \ref{coarse}, see also the discussion leading to equation
(\ref{cgsystem}). The largest eigenvalue is $0$ with three
corresponding pairs of left and right eigenvectors. All other
eigenvalues are negative and of order unity. For small $\varepsilon$
one can study the eigenvalues of ${\cal H} (\lambda)$ using
degenerate perturbation theory. The largest eigenvalue must be part
of the degenerate sector. Moreover, performing the leading order
perturbation theory leads to $\tilde{\cal H} (\lambda)$ as the
matrix whose eigenvalues are the leading order approximation for the
eigenvalues of ${\cal H} (\lambda)$ that belong to the degenerate
sector.

Therefore, the function $g^{\rm CG}(\lambda)$, obtained by computing
the largest eigenvalue of (\ref{calh}) approximately satisfies
$g^{\rm CG} (\lambda)=g^{\rm CG} (1-\lambda)$ (with corrections of
order $\varepsilon$). This holds not only for the largest eigenvalue
but to all the eigenvalues in the degenerate sector. This
perturbation theory for the Legendre transform of the large
deviation function can be considered as another justification for
the existence of an (approximate) fluctuation relation for the
coarse-grained entropy.

\begin{figure}[t]
\center{\includegraphics[scale=0.5]{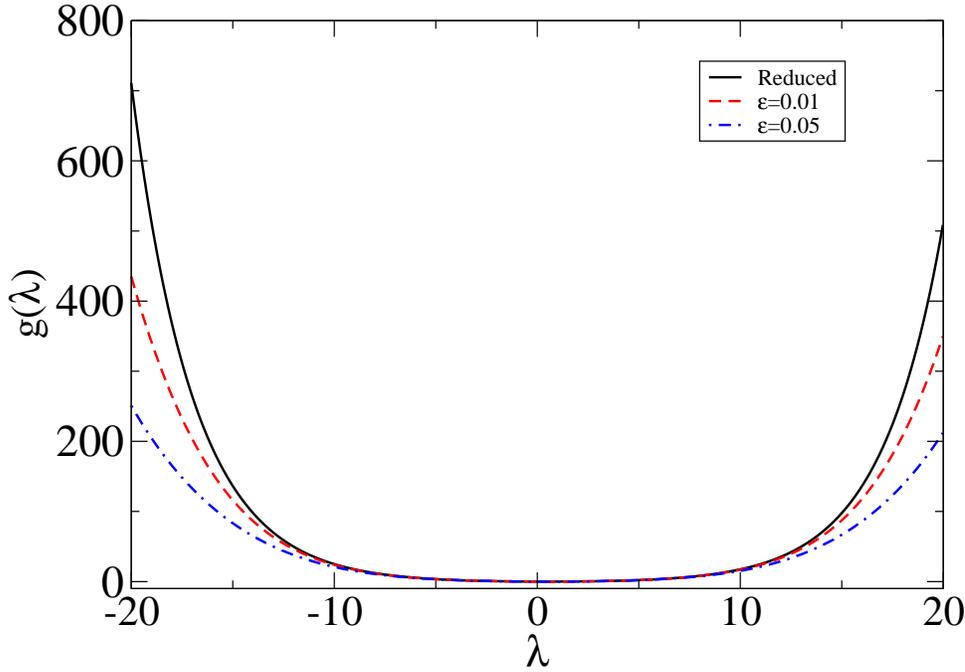}}
\caption{\label{ldvfig} The largest eigenvalue of i)  $\tilde{\cal
H}(\lambda)$, corresponding to the reduced system (solid line), ii)
${\cal H} (\lambda)$, pertaining to the coarse-grained dynamics,
with $\varepsilon=0.01$ (dashed), and with $\varepsilon=0.05$
(dashed-dotted). (All eigenvalues were divided by $\varepsilon$.)}
\end{figure}

We have calculated numerically the largest eigenvalue of ${\cal
H}(\lambda)$, as a function of $\lambda$ and compared it to the
largest eigenvalue obtained from $\tilde{\cal H}(\lambda)$. The
results are depicted in figure \ref{ldvfig}.
It is clear that the eigenvalue corresponding to the coarse-grained
system is a good approximation to the one obtained for the reduced
system. Moreover, the approximation improves for smaller
$\varepsilon$. The function $g^{\rm red} (\lambda)$ for the reduced
system vanishes at $\lambda=0$ and has the exact symmetry $g^{\rm
red}(\lambda)=g^{\rm red} (1-\lambda)$ as expected. The
corresponding functions obtained for the coarse-grained system start
to deviate from that function for larger values of $|\lambda|$. This
is related to the fact that the perturbation theory cannot be
uniformly valid. For instance, examining equation (\ref{calh}), it
is clear that, for any $\varepsilon$, one can find $\lambda^*$ so
that, say, $2^{\lambda^*} \varepsilon \simeq 1$. As a result, terms
which were assumed to be a small perturbation cease to be small.
This can be loosely interpreted as picking contributions from highly
unlikely trajectories, which may break our assumption of local
relaxation.

The Legendre transform of the function $g(\lambda)$ is the large
deviation function $f(q)$ which describes the asymptotic (with time)
behavior of the probability distribution of the entropy production.
The Legendre transform can be computed numerically. For
both systems studied here the maximum of $f(q)$ was found to be at
$\bar{q} \simeq 0.08$. This is the most likely value of the
coarse-grained entropy production rate in a long enough experiment,
in agreement with the entropy production rate calculated for the
steady state of the reduced system, $\bar{q}^{\rm red} \simeq
0.0833$. The fact that this value does not vanish indicates that the
system relaxes to
non-equilibrium steady state, in which detailed balance is violated.

\begin{figure}[t]
\center{\includegraphics[scale=0.5]{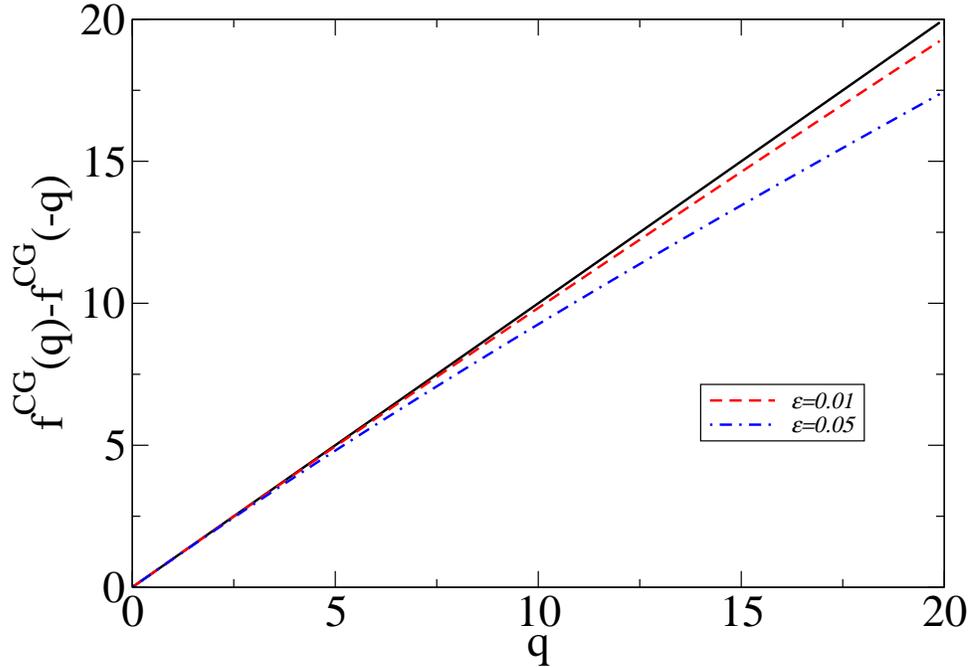}}
\caption{\label{prodfig} Plot of $f^{\rm CG}(q)-f^{\rm CG}(-q)$ as a
function of the measured coarse-grained entropy production rate $q$.
The solid line depicts the prediction of the fluctuation relation.}
\end{figure}

To verify that the coarse-grained entropy (approximately) follow the
Gallavotti-Cohen fluctuation relation we have calculated $f^{\rm
CG}(q)-f^{\rm CG}(-q)$ and compared it to the expected linear
behavior. The results are depicted in figure~\ref{prodfig}.
It is clear that fluctuation relation is a good approximation for
the coarse-grained entropy. The approximation improves for
smaller values of $\varepsilon$, as expected. Note that the
approximation deteriorates for larger values of entropy production
rates. One may speculate that this is a result of the increased
weight of trajectories that do not spend enough time in the
clusters to relax to the local stationary state.

The numerical results, presented in this section, were for a
coarse-grained entropy defined using a reduced system, which in turn
was obtained from the perturbation theory of section \ref{coarse}.
We have mentioned that there is another way to define a reduced
system, with the help of the steady state of the microscopic system.
The rates of this empirically-constructed reduced system differ from
the ones used so far by higher orders of $\varepsilon$. The
different rates will result in a different definition of the
coarse-grained entropy. One may wonder whether this coarse-grained
entropy will satisfy the fluctuation relation with much smaller
deviations. We have repeated the numerical calculation performed in
this section for this definition of coarse-grained entropy (with
$\varepsilon=0.05$, results not shown). The deviation from the
fluctuation relation were found to be similar to those presented in
figure~\ref{prodfig}.

\section{Conclusion}
\label{disc}

In this paper, we have studied stochastic jump processes which can
be coarse-grained. The coarse-grained dynamics can be viewed as the
dynamics of the system measured using a low resolution apparatus. A
perturbation theory, based on a separation of time scales, was used
to define a reduced system. The dynamics of this reduced system,
which is Markovian (and therefore satisfies an exact fluctuation relation),
approximates the coarse-grained system. This
reduced system then motivates the definition of a coarse-grained
entropy for the coarse-grained dynamics. This entropy was found to
approximately satisfy a fluctuation relation. Deviations from this
relation are more pronounced for large coarse-grained entropy
production rates. This deviation can be interpreted as a result of
the existence of internal degrees of freedom, which lead to
deviations from Markovian behavior of the coarse-grained entropy.

Let us consider an experiment whose goal is to measure fluctuation
relations. The considerations in this paper may be useful in the
interpretation of the results of such an experiment. If the results
exhibit deviations from the fluctuation relation one may suspect the
existence of unobserved degrees of freedom. (However, other
mechanisms, leading to deviations from the expected linear behavior,
exist~\cite{Cohen2007}.) Note that, even if there are internal
degrees of freedom, it is entirely possible that the entropy
production rates, needed for observing deviations from the
fluctuation relation, are so rare that such events will not be
measured during the experiment.

The results found in this paper lead to several new questions. For
instance, a better understanding of the deviation of the
coarse-grained entropy production from the fluctuation relation is
needed. The definition of the coarse-grained entropy used in this
paper was motivated by the definition applied to microscopic
systems. One may wonder whether it is possible to define another
coarse-grained entropy, using only quantities measured by the low
resolution apparatus, which will {\em exactly} satisfy the
fluctuation relation. It is also of interest to extend the
considerations, obtained here for stochastic systems, to
deterministic systems. Finally, we have seen that the existence of a
coarse-grained counterpart leads to a physically motivated
coarse-grained entropy exhibiting a fluctuation relation. One may
wonder whether the reverse is also true. Does the existence of a
quantity, which has some physical interpretation, and satisfies a
non-trivial fluctuation relation, suggest that the system can be
simplified in some way? Such questions are left for future work.

\section*{Acknowledgments}
We are grateful for start-up funds provided by the University of Maryland.

\end{document}